\documentclass[%
 reprint,
 amsmath,amssymb,
aps,
prb,
]{revtex4-2}

\usepackage{graphicx}
\usepackage{dcolumn}
\usepackage{bm}
\usepackage{hyperref}
\usepackage{xcolor}
\usepackage[mathlines]{lineno}
\usepackage{braket}
\usepackage{amsmath}

\begin{document}
\preprint{APS/123-QED}
\title{Quantum Fluctuations in the van der Walls material $\rm NiPS_3$}
\author{Paula Mellado}
\affiliation{
School of Engineering and Sciences, 
	Universidad Adolfo Ib{\'a}{\~n}ez,
	Santiago, Chile}
\author{Mauricio Sturla}
\affiliation{
Instituto de Física de Líquidos y Sistemas Biológicos (IFLYSIB),\\ UNLP-CONICET, La Plata, Argentina.}
\begin{abstract}
We present the magnetic excitation spectrum of the quantum magnet $\rm NiPS_3$ near the zig-zag ground state of a minimal honeycomb spin Hamiltonian that includes bilinear and biquadratic spin interactions. Our analysis, using a multi-boson generalized spin wave theory suited for spin S=1 systems,  revealed two normal modes at the linear level. The one at lower energy corresponds to a single magnon mode, consistent with results from spectroscopy experiments. Without single-ion anisotropy, this mode features a Goldstone mode at the corner of the Brillouin zone. When single ion anisotropy is introduced, the zig-zag phase's global U(1) invariance is broken, resulting in a gap. The higher energy mode corresponds to two-magnon fluctuations, which appear at the harmonic level in the generalized spin wave theory. This mode forms a gapped flat band due to bilinear spin interactions and becomes dispersive when biquadratic interactions are considered. The higher energy dispersion is related to quadrupolar fluctuations, which are feasible in magnets where the order parameter fluctuates in the SU(3) space. The spectrum analysis yielded quantum corrections to the order parameter and detected instabilities in the $\rm NiPS_3$ dipolar phases. Identifying the highest energy branch in experiments could provide insight into hidden nematic orders in $\rm NiPS_3$ and other van der Waals magnets.
\end{abstract}
\maketitle
\begin{figure}[htbp]
\includegraphics[width=\columnwidth]{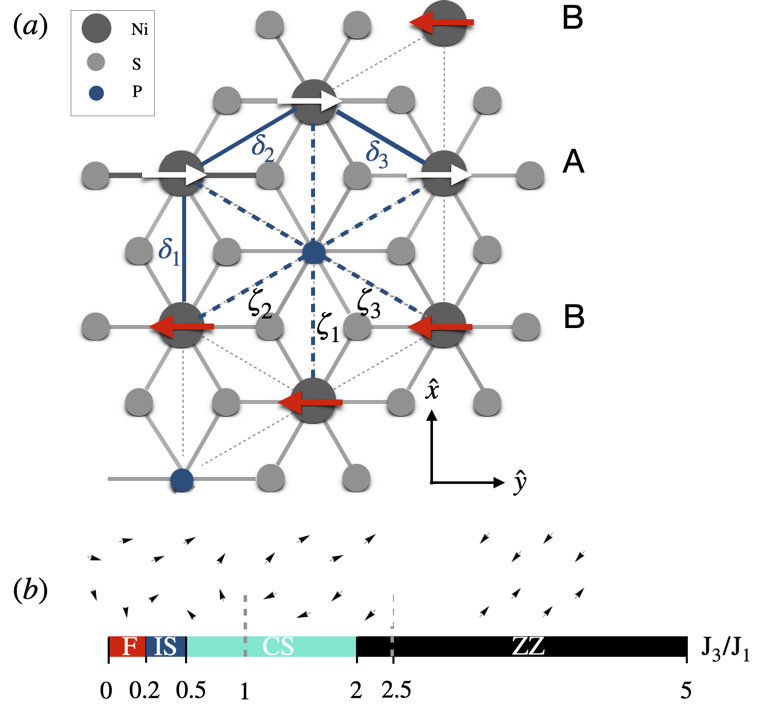}
\caption{(a) Sketch of  $\rm NiPS_3$ in the zig-zag  state. Red and white arrows denote spins in sublattices A and B, respectively. $\delta_1, \delta_2$ and $\delta_3$ are vectors joining nearest neighbor Ni atoms, $\zeta_1=2\delta_1, \zeta_2=2\delta_2$ and $\zeta_3=2\delta_3$ are the vectors joining third nearest neighbors. (b) Variational ground states as a function of $J_3$ for $A^z=0$ and $K_1=0.05J_1$ obtained in a cluster of 24 sites with open boundary conditions \cite{supp}. Dotted lines show that the boundaries between phases IS and CS and between CS and ZZ shift to larger $J_3$ when $\rm K_1=0$. Variational states ZZ, CS, and IS are depicted at the top of the phase diagram.}
\label{fig:f1}
\end{figure}
\begin{figure}[htbp]
\includegraphics[width=\columnwidth]{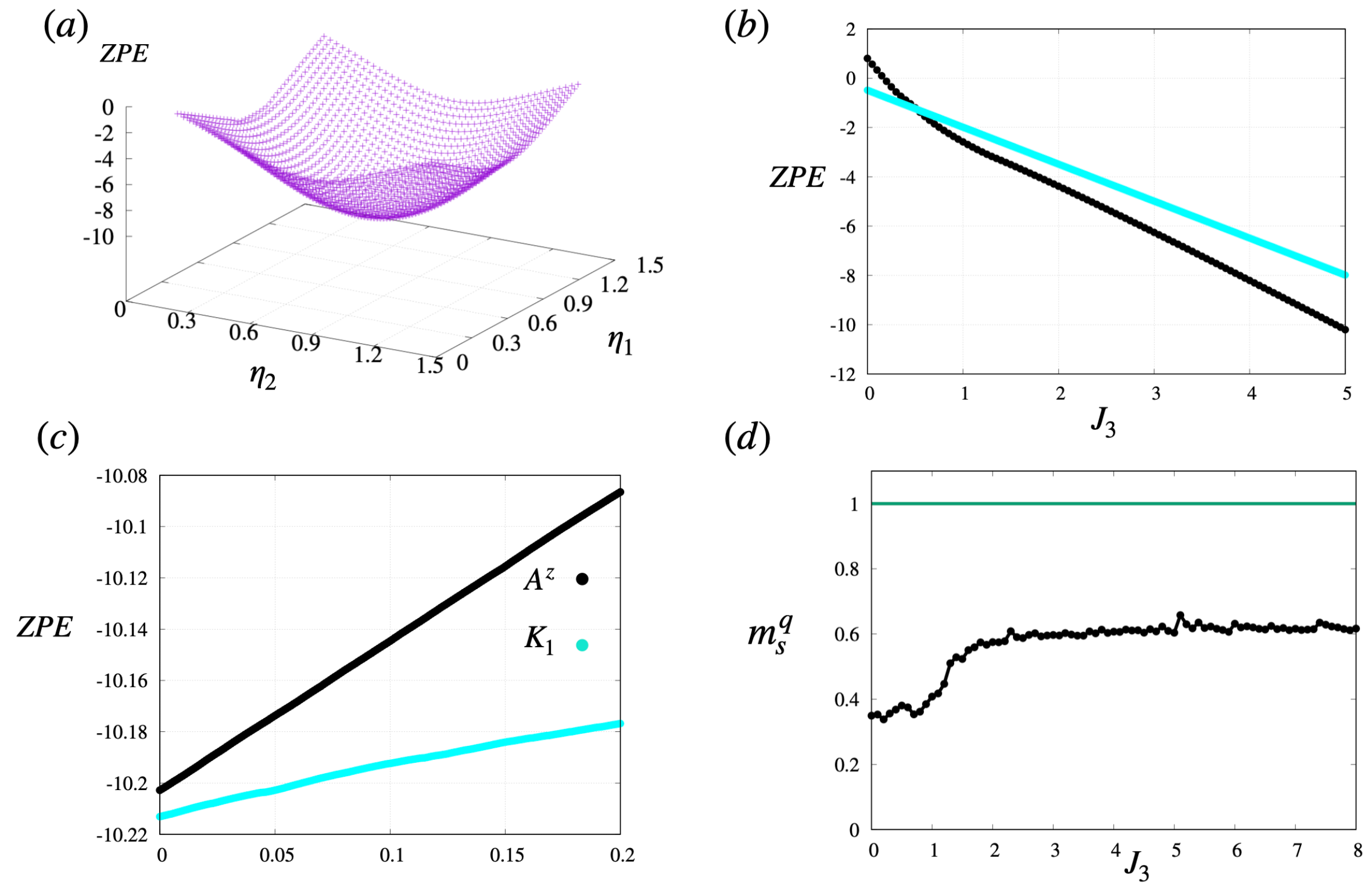}
\caption{(a) ZPE per site as a function of $\eta_1$ and $\eta_2$ at $K_1=0.05J_1, J_3=5J_1$. ZZ state corresponds to $\eta_1=\eta_2=\pi/4$ (b) ZPE as a function of $J_3$ for $K_1=0.05J_1$ and $A^z=0$ in black is compared to the ground state energy in cyan. (c) ZPE as a function of $K_1$ (cyan) and $A^z$ (black). (d) Staggered magnetization after 
quantum corrections $m_s^q$ as a function of $J_3$ for $ K_=0.05J_1, A^z=0$ compared to the ideal value $m_s=1$ in green.}
\label{fig:f3}
\end{figure}

\emph{Introduction.} 
Transition-metal thiophosphates \cite{burch2018magnetism}, are van der Waals compounds with reduced dimensionality that exhibit intriguing magnetic, electronic, and optical quantum effects \cite{seifert2022ultrafast,kim2018charge,gu2019ni,rosenblum1994two,rosenblum1999resonant,basnet2021highly, afanasiev2021controlling,belvin2021exciton,kang2020coherent,ergeccen2022magnetically,jiang2021recent,mattis2012theory,kim2023microscopic,wilson1969transition,basnet2022controlling, mak2019probing}. They realize a monoclinic space group $\rm C/2m$ where transition metal ions form a honeycomb lattice and are enclosed in octahedra formed by sulfur atoms. Magnetic susceptibility measurements have shown that the member of the family $\rm NiPS_3$ \cite{kim2021magnetic,lanccon2018magnetic,kertesz1984octahedral,hempel1981ground} has spin S=1, making it a suitable playground to explore the effects of quantum fluctuations and possible multipolar orders in an actual honeycomb spin system. Spectroscopy probes \cite{wildes2015magnetic, wildes2022magnetic,kim2019suppression,kim2021magnetic,kim2018charge,scheie2023spin,na2024direct} and DFT calculations \cite{chittari2016electronic, lane2020thickness,ushakov2013magnetism,wildes2015magnetic,wildes2022magnetic,hwangbo2021highly,lanccon2018magnetic} have shown that below the Neel temperature $\rm T_N=155$ K, the spins of Ni order magnetically and form a zig-zag antiferromagnetic pattern featured as double parallel ferromagnetic chains antiferromagnetically coupled within a layer (Fig.\ref{fig:f1}(a)). Spins in the zig-zag dipolar order are tilted in about $8^\circ$ out of it \cite{wildes2022magnetic,wildes2015magnetic}. High-resolution spectroscopy methods have proved spin dynamics in $\rm NiPS_3$ \cite{brec1986review, wildes2015magnetic}, and results from linear spin-wave theory using a Heisenberg Hamiltonian with single-ion anisotropies were used to estimate the magnetic exchange parameters and the nature of the anisotropy in $\rm NiPS_3$ samples \cite{olsen2021magnetic,kim2021magnetic,lanccon2018magnetic}. As a result, a ferromagnetic nearest-neighbor spin interaction $\rm J_1\sim 2.5$ meV and a dominant antiferromagnetic third-neighbor exchange coupling $\rm J_3\sim 13$ meV were found. In addition, easy plane anisotropy and a small uniaxial component were required to fit the experimental results, which led to two low-energy spin wave modes appearing in the spin-wave spectrum at the Brillouin zone center \cite{wildes2022magnetic}. The anisotropic Heisenberg Hamiltonian can reproduce the spin-wave energies but is at odds with the calculated neutron intensities \cite{chandrasekharan1994magnetism, kim2019suppression,lane2020thickness,wildes2015magnetic,kim2021magnetic, chittari2016electronic}. Furthermore, the moment magnitude of $\rm Ni^{2+}$ sites was found $\sim 1.05\mu_B$, smaller than the expected value $\sim 2\mu_B$ for the spin-only moment. This indicates that quantum fluctuations in $\rm NiPS_3$ play an essential role  \cite{scheie2023spin,wildes2015magnetic,wildes2022magnetic}.
\\
Recently, by studying the electron exchange mechanisms of a microscopic two-band Hubbard model for the Ni atoms in $\rm NiPS_3$ in the limit of strong Coulomb interactions, one of us derived the effective spin Hamiltonian of $\rm NiPS_3$ \cite{mellado2023spin}. In bulk samples, the crystal field at Ni sites causes the 3d orbitals to split into a lower energy triplet and a doublet, with the triplet fully occupied and the doublet half-filled. The spin interactions between Ni sites were derived by studying a superexchange process \cite{harrison2012electronic} mediated by sulfur orbitals \cite{mellado2023spin}. Calculations revealed a ferromagnetic bilinear nearest-neighbor interaction, a five times larger bilinear third-nearest neighbor antiferromagnetic interaction, and a ferromagnetic biquadratic spin coupling arising from microscopic grounds in  $\rm NiPS_3$.
\emph{Main results.} 
Here,  we used a generalized spin wave theory (GSWT) \cite{muniz2014generalized,penc2010spin} to examine quantum fluctuations and the stability of dipolar spin orders in $\rm NiPS_3$.
We found two double degenerate normal oscillation modes in the harmonic approximation. The softest branch, from the single magnon dispersion, is gapless at the $\rm {\bm k}=\mathcal{M}$ point of the hexagonal first Brillouin zone and aligns with a peak in the structure factor from spin-spin correlations, indicating a Goldstone mode \cite{auerbach2012interacting}. Including a single ion anisotropy opens a gap in the dispersion and softens the structure factor. The spin wave branch at higher energies is gapped, independent of single-ion anisotropies, and minimal at the corner of the Brillouin zone.  Its dispersive character is attributed to biquadratic interactions alone, and it is a fingerprint that could point to hidden nematic orders in future spectroscopic measurements.  By analyzing the spin wave dispersion, we identified conditions conducive to the instabilities of dipolar magnetic orders in  $\rm NiPS_3$. Comparison of the zero-point energy with ground state energy, in addition to calculations of quantum corrections to the order parameter, suggests that quantum fluctuations assist in stabilizing the magnetic order in $\rm NiPS_3$ by a mechanism akin to order by quantum disorder, which could explain the high Neel temperature of $\rm NiPS_3$.
\\
\emph{Biquadratic interactions}
In systems with spin S=1, dipolar and quadrupolar magnetic orders are possible \cite{chubukov1991spontaneous}. The quadrupolar operator is a tensor with five components ${\bm Q_i^{\alpha\beta}}=S^\alpha S^\beta+S^\beta S^\alpha-\frac{4}{3}\delta_{\alpha\beta}$. Using the identity $\rm{\bm Q_i}\cdot {\bm Q_j}=2({\bm S_i}\cdot {\bm S_j})^2+{\bm S_i}\cdot {\bm S_j}-2/3$ the biquadratic interaction can be written in terms of the bilinear interaction between quadrupolar operators, yielding the following bilinear-biquadratic effective spin Hamiltonian for the Ni atoms in $\rm NiPS_3$ \cite{mellado2023spin}:
\begin{eqnarray}
\mathcal{H}&=&-(J_1-\frac{K_1}{2})\sum_{<ij>}({\bf S}_i\cdot{\bf S}_j)
+J_3\sum_{(ik)}({\bf S}_i\cdot{\bf S}_k)\nonumber\\&& -\,\frac{K_1}{2}\sum_{<ij>}({\bf Q}_i\cdot{\bf Q}_j)-\sum_{i}\frac{4}{3}K_1  
\label{eq:HQ}
\end{eqnarray}
where $\rm J_1$ and $\rm J_3$ denote the first and third nearest neighbor bilinear spin exchange couplings, respectively, and $\rm K_1$ is the nearest neighbor biquadratic spin exchange. Spin couplings derived from the microscopic model were found in the range of experimental and DFT values in the proportion $J_3\sim 5J_1$, and a nearest neighbor biquadratic coupling not reported before was found to be in the range $K_1\in (0.01J_1-0.05J_1)$. To study the magnetic orders of Eq.\ref{eq:HQ} at T=0 \cite{mellado2023spin}, consider the trial ground state $ \ket{\psi}=\otimes_j\ket{\psi_j}$, an entanglement-free direct product of arbitrary wavefunctions with spin S=1 at each site $j$ \cite{lauchli2006quadrupolar,stoudenmire2009quadrupolar,ivanov2003effective,mattis2012theory}. The single spin state is the superposition of coherent states $\ket{\psi_j}=\sum_\alpha d_{j\alpha}\ket{\alpha}$, 
where $\bm{d}_j=\bm{u}_j+i \bm{v}_j$ is an arbitrary complex vector satisfying the normalization constraint $\bm{d}_j^*\cdot \bm{d}_j=1$ (phase fixing $\bm{u}\cdot \bm{v}=0$), and 
$\ket{\alpha}$ is the time-reversal invariant basis of the $\rm SU(3)$ fundamental representation for spin S=1 \cite{ivanov2003effective, batista2002unveiling} with $\alpha=x,y,z$ on every site. The spin operator in this basis is $S^\alpha=-i\sum_{\beta\gamma}\epsilon_{\alpha\beta\gamma}\ket{\beta}\bra{\gamma}$ and the magnetization of the system is defined at each site as \cite{stoudenmire2009quadrupolar}
$M=\sum_j\bra{\psi_j}{\bf S}_j\ket{\psi_j}=-i\sum_j\bm{d}_j^*\times \bm{d}_j$.
In the relevant space of parameters for bulk $\rm NiPS_3$ \cite{lanccon2018magnetic} samples, the variational ground state of Eq.\ref{eq:HQ} corresponds to the zig-zag Neel magnetic order illustrated in Fig.\ref{fig:f1}(a) \cite{wildes2022magnetic,scheie2023spin,mellado2023spin,lanccon2018magnetic,lane2020thickness}, where spins tilt out of the lattice plane. When $K_1$ takes values in the range expected in actual samples ($\rm K_1\leq 0.1J_1$), the zig-zag state (ZZ) is the ground state of the system as long as $J_3>2J1$ as shown in Fig.\ref{fig:f1}(b). In the range $0.5J1\lesssim J_3\lesssim 2J1$, spins order in a commensurate spiral phase (CS) \cite{mellado2023spin}. CS evolves into a twisted state resembling an incommensurate spiral (IS) in the $0.2J1\lesssim J_3\leq 0.4J1$ range. Finally, at $J_3\lesssim 0.2J1$, the system settles in a ferromagnetic order (F). These variational states prevail when $K_1=0$, but boundaries between ZZ/CS and CS/IS shift to larger values of $J_3$ as indicated by the vertical dotted lines in Fig.\ref{fig:f1}(b). The biquadratic coupling primarily forces nearest neighbor spins to be parallel.
\\
In previous studies, phases ZZ and F were observed in the classical XY and Heisenberg honeycomb $J_1-J_3$ models by \cite{rastelli1979non} within the respective ranges $J_3 \gtrsim 0.4 J_1$ and $J_3 \lesssim 0.2 J_1$. In both scenarios, a helical phase was identified in the intermediate range, $0.2 J_1 \lesssim J_3 \lesssim 0.4 J_1$. Additionally, this twisted phase was also discovered in the classical limit of the quantum counterpart $S=\frac{1}{2}$ by \cite{jiang2023quantum} using a long-scale renormalization group combined with an augmented spin wave theory (SWT). In this work \cite{jiang2023quantum}, a ferromagnetic phase was found in the quantum XXZ model in the limits of XY and Heisenberg spins in the respective ranges $J_3  \lesssim 0.28 J_1$ and $J_3  \lesssim 0.25 J_1$. In the XY case, the ZZ phase occurred at $J_3 \gtrsim 0.5 J_1$ while an Ising zig-zag phase was reported in the range $0.28 J_1\lesssim J_3 \lesssim 0.55 J_1$. In the Heisenberg model, ZZ realized at $J_3 \gtrsim 0.26 J_1$ while a double zig-zag order was reported in the range $0.23 J_1\lesssim J_3 \lesssim 0.28 J_1$ \cite{jiang2023quantum}. 
\\
In $\rm NiPS_3$, spins are in a triplet state, and the orbital angular momentum is a singlet; thus a small spin-orbit interaction is expected \cite{lanccon2018magnetic,kim2021magnetic, auerbach2012interacting}. To determine its effect, in \cite{mellado2023spin} the term $\rm A^z\sum_{j}(S_j^z)^2$ was added to Eq.\ref{eq:HQ}, where $A^z$ plays the role of the anisotropic coupling \cite{autieri2022limited,olsen2021magnetic,lane2020thickness}. Besides enforcing spins to settle close to the $x-y$ plane, we found that like $K_1$, $A^z$ shifts the variational boundaries of Fig.\ref{fig:f1}(b) toward smaller $J_3$ values \cite{mellado2023spin}. Therefore, $K_1$ and $A^z$ stabilize variational phases toward smaller values of $J_3$.
 \begin{figure}
\includegraphics[width=\columnwidth]{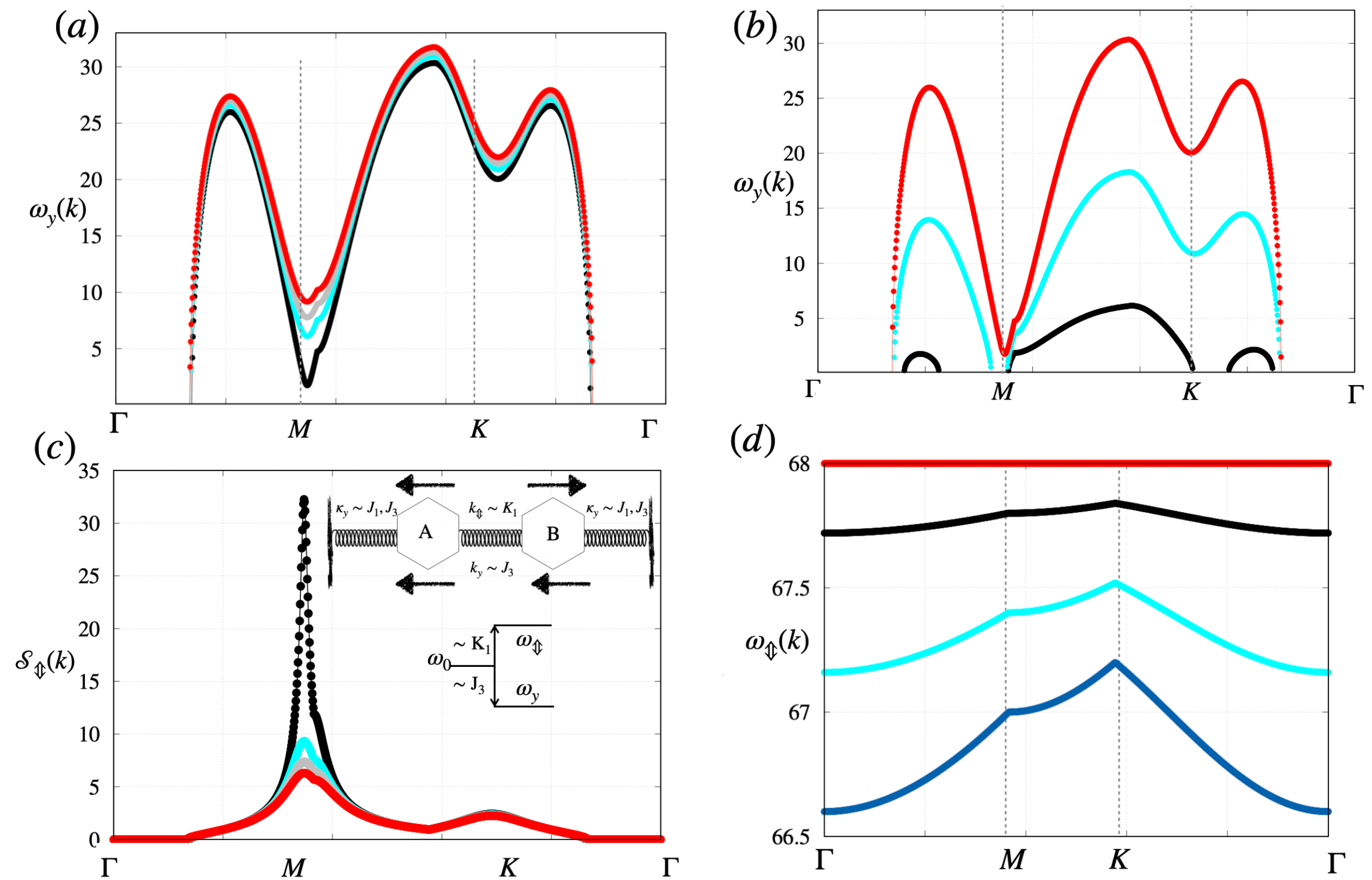}
\caption{(a) $\omega_y(k)$ with $\rm K_1=0.05J_1, J_3=5J_1$ and $A^z=0.2J_1,0.5J_1,0.7J_1,0.9J_1$ in black, cyan, grey and red respectively, (b) $\omega_y(k)$ with $\rm A^z=0.2, \rm K_1=0.05J_1$ and $\rm J_3=J_1, 3J_1, 5J_1$ in black, cyan and red, respectively. (c) $\mathcal{S}_\Updownarrow(k)$ with $\rm A^z=0.2J_1, 0.5J_1, 0.7J_1, 0.9J_1$ in black, cyan, blue and red respectively.  Inset: Illustration of the effect of GSWT in splitting the traditional spin wave dispersion $\omega_0$ and the main roles played by $J_3$ and $K_1$ in each mode. (d) $\omega_\Updownarrow(k)$ with $\rm K_1=0J_1, 0.01J_1, 0.03J_1, 0.05J_1$ in red, black, cyan and blue respectively.}
\label{fig:f2}
\end{figure}
\\
\emph{Quantum fluctuations.} 
Quantum fluctuations in S=1 spin systems are accurately described by extending the local SU(2)-group of local spin rotations, used in the traditional spin wave theory (SWT), to the SU(3) group of local unitary transformations as shown in \cite{muniz2014generalized}. This approach has been denoted generalized spin wave theory, or multi-boson SWT \cite{muniz2014generalized}. In this framework, the total number of Schwinger bosons (SB) in every site is constrained $n_b=\sum_\alpha a_\alpha^{\dagger}(j)a_\alpha(j)$  \cite{lauchli2006quadrupolar,muniz2014generalized,papanicolaou1988unusual,remund2022semi}. $n_b$=1 describes S=1 spins \cite{lauchli2006quadrupolar,muniz2014generalized,penc2010spin}. Construction of the SU(3) generators is based on three independent pairs of destruction and creation operators \cite{penc2010spin}. In terms of these SB, spin, and quadrupole operators are: $S_j^\alpha=-i\epsilon_{\alpha\beta\gamma}a_{\beta}^{\dagger}(j) a_{\gamma}(j), \quad Q_{j}^{\alpha\beta}=\frac{2n_b}{3}\delta_{\alpha\beta}-a_{\alpha}^{\dagger}(j)a_{\beta}(j)-a_{\beta}^{\dagger}(j)a_{\alpha}(j)$. An ordered state of SU(3) spin-coherent states corresponds to condensing the appropriate boson in each sublattice. The remaining ones play the role of the Holstein-Primakoff bosons. The information about which boson to condense is obtained from the variational wave function and quantum fluctuations are considered via a $1/n_b$ expansion. This gives a generalized spin wave theory, suitable for S=1 systems, where both spin and quadrupolar waves are accounted for \cite{lauchli2006quadrupolar}. 
\\
To construct the variational uniform ZZ state, consider identical on-site wave functions for all spins on the same sublattice. Single-site wavefunctions in sublattices A and B are $\ket{\psi}^{A}=\cos{\eta_1}\ket{x}+i\sin{\eta_1}(\sin{\alpha_1}\ket{y}+\cos{\alpha_1}\ket{z}), \quad \ket{\psi}^{B}=\cos{\eta_2}\ket{x}-i\sin{\eta_2}(\sin{\alpha_2}\ket{y}+\cos{\alpha_2}\ket{z})$. Such normalized wavefunctions satisfy that the magnetic moment of sublattices A and B point in opposite directions in the plane $y-z$ and the length of spins is $S^{A}=\sin{(2\eta_1)}^2$ and $S^{B}= \sin{(2\eta_2)}^2$ respectively. Minimization of the expectation value of Hamiltonian Eq.\ref{eq:HQ} respect to $\eta_1$ and $\eta_2$ at $J_1=1, J_3=5J_1, K_1=0.05J_1 $ yields $\eta_1=\eta_2=\pi/4$ and $\alpha_1=-\alpha_2$. These states are fully magnetic states with no (nontrivial) quadrupolar order and $u^2=v^2=\frac{1}{2}$. Without loss of generality, we set $\eta=\pi/4$ and $\alpha=0$ that corresponds to the zig-zag Neel state $\ket{0}^{A(B)}=\frac{\sqrt{2}}{2}(\ket{x} \pm i\ket{z})$ with fully polarized spin vectors along the axis $\pm\hat{y}$, Fig.\ref{fig:f1}(a). A global rotation in $\pi/4$ of $a_x$ and $a_z$ around $\hat{y}$ gives rise to rotated SB $a_\uparrow^\dagger=\frac{(a_x^\dagger+i a_{z \dagger})}{\sqrt{2}},\quad  a_\downarrow=\frac{(a_x+i a_z)}{\sqrt{2}}$ that are replaced in expressions for spins and quadrupolar operators, as detailed in the Supplemental Material \cite{supp}.  The ZZ ordered state is set by condensing $a_\uparrow$  in sublattice A and $a_\downarrow$ in sublattice B. Introducing propagating states on each sublattice via a Fourier transformation $a_{\mu i}^\dagger=\sqrt{\frac{2}{L}}\sum_{\bm k}\exp{(-i{\bm k\cdot \bm R_i})}a_{\mu{\bm k}}^\dagger$
where $L$ is the total number of sites in the honeycomb lattice, and $\bm k$ are vectors in its reduced Brillouin zone (RBZ), Eq.\ref{eq:HQ} of order $\mathcal{O}(n_b^2)$ becomes
\begin{eqnarray}
H^{(2)}=H_0+H_y(a_y,b_y)+H_{\Updownarrow}(a_{\downarrow},b_{\uparrow})
\label{eq:hsw}
\end{eqnarray}
where SB of sublattice A and B are denoted by $a$ and $b$ respectively and  $H_0=-2 L\left[J_1-\frac{K1}{2}+3J_3\right]$. $H_y$ and $H_{\Updownarrow}$ can be diagonalized independently 
because there aren't coupled terms of the type $a_ya_\downarrow$. A bosonic Bogoliubov transformation $a^\dagger=u d_1^\dagger+vd_2,\quad
b^\dagger=u d_2^\dagger+vd_1$, eliminates anomalous pair creation terms in $H_y$ and $H_{\Updownarrow}$, and in their diagonal form they become
\begin{eqnarray} 
H_{yk}&=\sum_k\omega_{yk}\left[d_{1y}^\dagger d_{1y}+d_{2y}^\dagger d_{2y}\right]+\omega_{yk}-\epsilon_{yk} \\ 
H_{\Updownarrow k}&=\sum_k\omega_{\Updownarrow k}\left[d_{1\downarrow}^\dagger d_{1\downarrow}+d_{2\uparrow}^\dagger d_{2\uparrow}\right]+\omega_{\Updownarrow k}-\epsilon_{\Updownarrow k}
\end{eqnarray}
where  $\omega_{y k}=\sqrt{\epsilon_{y k}^2-\lambda_{y k}^2}$, $\omega_{\Updownarrow k}=\sqrt{\epsilon_{\Updownarrow k}^2-\lambda_{\Updownarrow k}^2}$ 
and \begin{align} 
\epsilon_{y k}&=-4\left[J_1-K_1\right]\xi^{\parallel}_k+2\left[3J_3+J_1-K_1\right]\label{eq:epy}\\ 
\lambda_{y k}&=4\left[J_1-K_1\right]\xi^{\perp}_k-4J_3\beta^{\perp}_k\label{eq:lay}\\
\epsilon_{\Updownarrow k}&=-8K_1\xi^{\parallel}_k+12\left[\frac{2}{3}J_1-K_1+J_3\right]\label{eq:epa}\\ 
\lambda_{\Updownarrow k}&=-4K_1\xi^{\perp}_k\label{eq:laa}
\end{align} Geometric factors, $\xi^{\parallel}_k=\cos{k\delta_2}+\cos{k\delta_3}$, $\xi^{\perp}_k=\cos{k\delta_1}$, $\beta^{\perp}_k=\cos{k\zeta_1}+\cos{k\zeta_2}+\cos{k\zeta_3}$ couple respectively nearest neighbor spins in the same sublattice,  nearest neighbors in different sublattices and third neighbor spins in different sublattices, as shown in Fig.\ref{fig:f1}(a).
Branches $\omega_{y k}$ and $\omega_{\Updownarrow k}$ are two-fold degenerate transverse spin waves: the normal oscillation frequencies of spins in sublattices A and B. The softest mode $\omega_y$, Fig.\ref{fig:f2}(a),(b) corresponds to spins in the two sublattices precessing in-phase about the classical axes $\pm\hat{y}$. $\omega_y$ originates fluctuations in the dipolar magnetic order of $\rm NiPS_3$ \cite{supp}. It is the dispersion associated with single-magnon frequencies.
GSWT yields a second mode at higher energies $\omega_\Updownarrow$, hidden in the usual Holstein-Primakoff formalism.
$\omega_\Updownarrow$ accounts for the out-of-phase oscillations of spins in the two sublattices and corresponds to two-magnon modes due to quadrupolar fluctuations in $\rm NiPS_3$. When $K_1=0$,  $\omega_\Updownarrow$ is a gapped dispersionless flat mode. Once $K_1$  is turned on, $\omega_\Updownarrow$ becomes dispersive by coupling out-of-phase quantum fluctuations in both sublattices. The two magnons are bound by $K_1$, and dispersion $\omega_\Updownarrow$ corresponds to quadrupolar fluctuations. $\omega_y$ and $\omega_\Updownarrow$ can be understood as follows: $\beta^\perp$ couples the sublattices by connecting sites in sublattice A with three sites in sublattice B that are related by $C_3$ symmetry; it is a symmetric sublattice coupling (see Fig.\ref{fig:f1}(a)). On the contrary, $\xi^{\perp}$ breaks the $C_3$ point symmetry of the honeycomb lattice by coupling sites in different sublattices along one of the three possible axes. As $\rm K_1\ll J_3, J_1$, low energy fluctuations $\omega_y$ are dominated by bilinear terms $J_1$ and $J_3$ respectively, Eqs.\ref{eq:epy},\ref{eq:lay}. On the other hand, out-of-phase antisymmetric inter-lattice couplings from anomalous boson contributions are only caused by $K_1$. This is illustrated in the inset of Fig.\ref{fig:f2}(c). In a linear SWT,  single magnon states oscillate with the same frequency $\omega_0$ in both sublattices. GSWT couples transverse fluctuations between sublattices through the $a_\Updownarrow$ SB. As a result, $\omega_0$ splits, as illustrated in Fig.\ref{fig:f2}(c). When biquadratic interactions are absent, $\omega_\Updownarrow$ becomes a flat band whose width $\Delta_\Updownarrow=4(2J_1+3J_3)$, Fig.\ref{fig:f2}(d) is associated with the cost of local rotations of spins in both sublattices.
\\
$\omega_y$ is minimal and gapless at the $\mathcal{M}$ point of the RBZ.
The zero mode from the U(1) symmetry associated with continuous spin rotations in the $\hat{y}-\hat{z}$ plane stands regardless of the value of $K_1$. As $K_1$ gets larger, it shifts barely $\omega_{yk}$ since it slightly rescales the nearest neighbor exchange 
$J_1\longrightarrow J_1-K_1$. 
\\
Inclusion of an single-ion anisotropy \cite{wildes2015magnetic,wildes2022magnetic,lanccon2018magnetic,lane2020thickness,olsen2021magnetic,scheie2023spin} in the multi-boson language $(S^z_k)^2=\frac{n_b}{2}\left(a_{yk}^\dagger a_{yk}+n_b\right)$ adds a constant to $\epsilon_{y k}$ becoming $\bar{\epsilon}_{y k}=\epsilon_{y k}+2A^z$ opening a gap at the $\mathcal{M}$ point: $\bar{\omega}_y(\mathcal{M})=\sqrt{4A^z (A^z-2J_1+6J_3+2 K_1)}$ as shown in Fig.\ref{fig:f2}(a).  Consequently, single-ion anisotropy and biquadratic interaction leave different signatures in the spin wave dispersion. 
\\
The imaginary part of the spin-spin correlation functions,
$\mathcal{S}_{\Updownarrow}(k,\omega)= n_b\delta(\omega-\bar{\omega}_{y k})\sqrt{\frac{\bar{\epsilon}_{y k}+\lambda_{y k}}{\bar{\epsilon}_{y k}-\lambda_{y k}}}$ \cite{penc2010spin,lauchli2006quadrupolar} arises from bosons $a_y$ in $H^{(2)}$. Analogously, $a_\uparrow$ and $a_\downarrow$ yield the spin-spin correlation function 
$\mathcal{S}_{y}(k,\omega)$. Structure factors are found integrating the frequency out. 
Figure \ref{fig:f2}(c) shows single-magnon spin fluctuations $\mathcal{S}_\Updownarrow(k)$ as a function of $A^z$ for $K_1=0.05J_1, J_3=5J_1$. The softening of $\omega_y(\mathcal{M})$ coincides with peaks in $\mathcal{S}_\Updownarrow(k)$, indicating that the zig-zag phase is close to another phase, possibly CS, as shown in Fig.\ref{fig:f1}(b). Goldstone theorem \cite{auerbach2012interacting,fazekas1999lecture} indicates that the peak of $\mathcal{S}_\Updownarrow(\mathcal{M})$ at $A^z=0$ is rooted in the existence of a Goldstone mode associated to the global U(1) invariance at $A^z=0$. 
\\
\emph{Stability of dipolar phases and order by quantum fluctuations.}
The stability of F and ZZ dipolar ordering requires that the dispersion modes have semi-positive defined frequencies at the $\Gamma$ and $\mathcal{M}$ points, respectively. Instability of the ferromagnetic state occurs for $\bar{\omega}_y(\Gamma )=0$ at $J_3^c=\frac{1}{9}[5(J_1-K_1)-A^z]$.  Evaluated at $K_1=0.05J_1$ and $A^z=0.2J_1$ we find that the limit of the F phase in $\rm NiPS_3$ occurs at a $J_3^c\sim 0.5$, larger than the variational calculation, Fig.\ref{fig:f1}b. This is expected as the variational phase diagram of Fig.\ref{fig:f1}b is computed using Eq.\ref{eq:HQ} (the system has Heisenberg symmetry), but the GSWT analysis is performed around a uniform ZZ phase with spins in sublattice A and B pointing along easy axes $\pm\hat{y}$ (the system has a U(1) symmetry). Instability of the zig-zag state occurs for $\bar{\omega}_y(\mathcal{M})=0$. This condition is met as long as $A^z=0$.
\\
Close to the $\mathcal{M}$ point, and for small values of $K_1$, $\bar{\omega}_y(k\longrightarrow \mathcal{M})\sim v_y|\bm{k}-\mathcal{M}|$, with magnon velocity $v_y\sim \frac{2\sqrt{3}(J_1-K_1)}{A^z}\sqrt{A^z(A^z-2J_1+6J_3+2K_1)}$. The structure factor scales as $\mathcal{S}_{\Updownarrow}(k\longrightarrow \mathcal{M})\sim \chi_y v_y|\bm{k}-\mathcal{M}|$,  where $\chi_y=-\frac{1}{2A^z}$ is the mean field susceptibility.
The dispersion associated with the two-magnon fluctuations at $\mathcal{M}$ scales linear with $\bm k$ too, with the velocity of two-magnon mode proportional to biquadratic interactions $v_\Updownarrow(\mathcal{M})\sim 4K_1\sqrt{3}$. The mean field susceptibility in this case becomes $\chi_\Updownarrow=-\frac{\sqrt{3}}{24(2J_1+3J_3)}$. In the long wave limit, we find that $\omega_\Updownarrow \sim \frac{k^2}{2 m_\Updownarrow}$ have a quadratic dispersion, where $m_\Updownarrow \sim 1/4K_1$ is the effective mass of the two-magnon excitation. 
\\
The comparison of the zero point energy due to quantum fluctuations $ZPE=\omega_{yk}+\omega_{\Updownarrow k}-[\epsilon_{yk}+\epsilon_{\Updownarrow k}]$ \cite{lauchli2006quadrupolar,auerbach2012interacting} with the energy of the ground state $E_0=\bra{0}\mathcal{H}\ket{0}$ allows to weight the relevance of quantum fluctuations in $\rm NiPS_3$.  In Figure \ref{fig:f3}(a) we show that $ZPE$ of  $\ket{\psi}^{\rm A,B}$ at $\alpha_1=-\alpha_2$ is minimum when the system is in the zig-zag state, $\eta_1=\eta_2=\pi/4$,  in the relevant range of spin couplings for $\rm NiPS_3$,  $J_3=5J_1$ and $K_1=0.05J_1$. Figures \ref{fig:f3}(b,c) show that while biquadratic and single-ion anisotropy terms have a minor effect suppressing quantum fluctuations on the zig-zag state of $\rm NiPS_3$, $J_3$ plays a major role, reducing the ground state energy $E_0$ of the zig-zag state and thus stabilizing this phase.  Sublattice magnetization in the zig-zag state are $S^A=\frac{L}{2}-\sum_{m,k} a_{mk}^\dagger a_{mk}$, and $S^B=-\frac{L}{2}+\sum_{m,k} b_{mk}^\dagger b_{mk}$. The order parameter of the ZZ phase is the staggered magnetization $m_s$. Quantum fluctuations reduce $m_s$
from its classical value to $m_s^q\sim 1-\frac{1}{2L}\sum_k\left[\frac{1}{\sqrt{1-\left(\lambda_{yk}/\epsilon_{yk}\right)^2}}+1\right]$ \cite{muniz2014generalized, auerbach2012interacting, lauchli2006quadrupolar,scheie2023spin}. Figure \ref{fig:f3}(d) shows $m_s^q$ as a function of $J_3$. We note that for $J_3<J_1$, (at $K_1=0.05J_1$) $m_s^q$ depicts a sudden drop. Finding a divergent spin reduction shows that the long-range zig-zag order is unstable for this range of $J_3$ in agreement with the phase diagram of Fig.\ref{fig:f1}(b). This is not the case for the classical system nor the $S=1/2$ case for which the zig-zag order has been shown to persist for $J_3< J_1$ \cite{rastelli1979non, jiang2023quantum}.
\begin{figure}[htbp]
\centering\includegraphics[width=0.9\columnwidth]{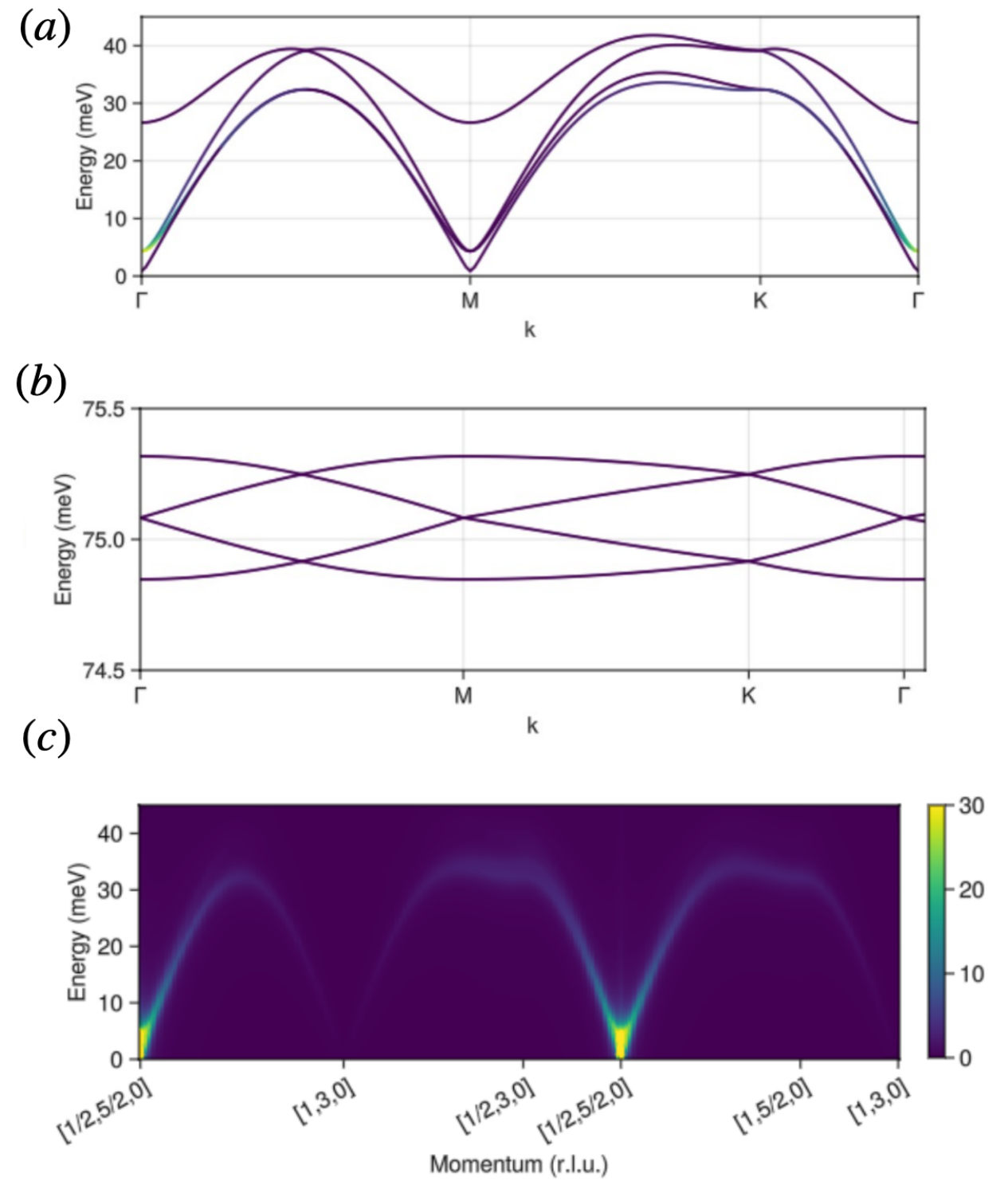}
\caption{Landau-Lifshitz approach as implemented in SU(N)NY. (a) $\omega_y(k)$ and (b) $\omega_\Updownarrow(k)$, for $J_3 = 5J_1, K_1=0.05J_1, A^z =0.1J_1, A^y=-0.005J_1$. (c) Inelastic neutron scattering spectrum simulated by  SU(N)NY, for $J_3 = 5J_1, K_1=0.05J_1, A^z =0.1J_1, A^y=-0.005J_1$.}
\label{fig:f4}
\end{figure}
\\
\emph{ Numerical Results.} To compare directly with neutron scattering experiments \cite{scheie2023spin} we used the generalized spin wave package SU(N)NY software suite \cite{SunnyPackage} calculating $SU(3)$ ($S=1$) Landau-Lifshitz dynamics  \cite{Dahlbom2022Sunny} on a $[1-1 0; 1 1 0; 0 0 1]$ supercell. Figs. \ref{fig:f4} (a-b) show the band structure for (a) $\omega_y(k)$ and (b) $\omega_\Updownarrow(k)$ when  $J_3 = 5J_1, K_1=0.05J_1, A^z =0.1J_1, A^y=-0.005J_1$, where the weakly dispersive character of $\omega_\Updownarrow(k)$  is apparent. Fig \ref{fig:f4}(c) shows the calculated inelastic neutron scattering spectrum for the same set of parameters. The spectrum thus obtained is in good accordance with the experimental results of ref. \cite{scheie2023spin}.  The distinctive band $\omega_\Updownarrow(k)$ is not visible in the scattering data along this specific high-symmetry direction.  Further experiments probing the dispersion of this upper band along other directions could give a definite and quantitative answer regarding the presence of biquadratic interactions.

\emph{Discussion}  
In $\rm NiPS_3$, a honeycomb quantum spin system, not only dipolar magnetic orders but also quadrupolar states could be observed. This is because the $\rm Ni^{2+}$ ions in $\rm NiPS_3$ have spin $S=1$. The low energy excitations in this system are waves of the local order parameter that fluctuate in the SU(3) space of unitary transformations of the local spin. Considering this, we used a generalized version of the spin wave theory to analyze quantum fluctuations around the ZZ state in $\rm NiPS_3$. 
\\
The single-magnon mode at the lowest energy features a Goldstone mode at the corner of the RBZ, which disappears with finite values of a single-ion anisotropy, which breaks the U(1) global symmetry. Consequently, without anisotropy, the ZZ phase is unstable relative to other possible magnetic orders. Quadrupolar fluctuations correspond to the highest energy mode at the harmonic level in GSWT. We found that the dispersive character of this two-magnon state relies on biquadratic interactions, while bilinear spin interactions dominate the bandwidth. 
\\
Close to the $\mathcal{M}$ point, the single-magnon dispersion scales linearly with $\bm k$ with a magnon velocity that depends on all spin couplings and decays with anisotropy as $\sim (A^z)^{-1/2}$. The mean field susceptibility is negative and decreases as $\sim (A^z)^{-1}$.
\\
GSWT predicts that the zig-zag phase is unstable in the absence of $A^z$, indicating that quantum fluctuations help to stabilize the zig-zag order in samples of $\rm NiPS_3$. This scenario is consistent with the zero point energy being smaller than the ground state energy at $J_3>J_1$. Quantum effects are also evident in reducing the staggered magnetization from its classical value. The further reduction when  $J_3\lesssim J_1$ indicates that the long-range zig-zag order is unstable within this range of $J_3$.
\\
We conclude our work with three messages. 1) Besides the lowest energy mode already found in spectroscopic experiments, another mode associated with quadrupolar fluctuations is part of the spectrum of the spin fluctuations in $\rm NiPS_3$ at frequencies twice as large and remains to be found. 2) If the high energy mode is dispersive in experiments, this is a sign that biquadratic interactions in the system are sizable in the material \cite{sato2009nmr}. 3) The zig-zag phase is close to a spiral phase, which should coexist with some quadrupolar state. One possible way to engineer this scenario is to modify $J_3$ using chemical substitution \cite{autieri2022limited}, which has proven an effective tool for tuning exchange constants.  This could shed light on the nature of the dipolar and quadrupolar orders in bilayers and monolayers of $\rm NiPS_3$.
\\

\emph{Acknowledgments.}  P.M. Thanks, Santiago Griguera, K. Penc,  A. Scheie, A. Chernyshev, N. Drichko, D. A. Tennant, and Sylvain Cappone for useful discussions. This work was supported in part by Fondecyt under Grant No. 1210083 (P.M.) and was performed in part at the Aspen Center for Physics, which is supported by a grant from the Simons Foundation (1161654, Troyer) (P.M.). 
%

\end{document}